# Carrier localization and electronic phase separation in a doped spin-orbit driven Mott phase in $Sr_3(Ir_{1-x}Ru_x)_2O_7$

Chetan Dhital,[1] Tom Hogan,[1] Wenwen Zhou,[1] Xiang Chen,[1] Zhensong Ren,[1] Mani Pokharel,[1] Yoshinori Okada[1], M. Heine[1], Wei Tian,[2] Z. Yamani,[3] C. Opeil,[1] J. S. Helton,[4] J. W. Lynn,[4] Ziqiang Wang,[1] Vidya Madhavan,[1] and Stephen D. Wilson[1,*]

[1]*Department of Physics, Boston College, Chestnut Hill, Massachusetts 02467, USA*

[2]*Neutron Scattering Science Division, Oak Ridge National Laboratory, Oak Ridge, Tennessee 37831-6393, USA*

[3]*Chalk River Laboratories, Canadian Neutron Beam Centre, National Research Council, Chalk River, Ontario, Canada K0J 1P0*

[4]*NIST Center for Neutron Research, National Institute of Standards and Technology, Gaithersburg, Maryland 20899-6102, USA*

* *Address correspondence to stephen.wilson@bc.edu*

**Interest in many strongly spin-orbit coupled *5d*-transition metal oxide insulators stems from mapping their electronic structures to a $J_{eff}$=1/2 Mott phase. One of the hopes is to establish their Mott parent states and explore these systems' potential of realizing novel electronic states upon carrier doping. However, once doped, little is understood regarding the role of their reduced Coulomb interaction *U* relative to their strongly correlated 3d-electron cousins. Here we show that, upon hole-doping a candidate $J_{eff}$=1/2 Mott insulator, carriers remain localized within a nanoscale phase separated ground state. A percolative metal-insulator transition occurs with interplay between localized and itinerant regions, stabilizing an antiferromagnetic metallic phase beyond the critical region. Our results demonstrate a surprising parallel between doped 5d- and 3d-electron Mott systems and suggest either through the near-degeneracy of nearby electronic phases or direct carrier localization that *U* is essential to the carrier response of this doped spin-orbit Mott insulator.**

$Iridium^{4+}$ ions with a half-filled 5d shell in a cubic octahedral oxygen coordination occupy a unique region in relative energy scales: one where a model of crystal field splitting combined with strong spin-orbit coupling (SOC) breaks the five-fold degeneracy of electronic states into fully occupied $J_{eff}$=3/2 and half-filled $J_{eff}$=1/2 bands[1,2]. The resulting bandwidth-narrowed $J_{eff}$=1/2 states allow the relatively modest *U* (~1.5-2 eV) inherent to these 5d-transition metal elements [3] to split the band and generate a charge-gap. A SOC-assisted Mott phase results, allowing an unexpected manifestation of correlation-driven physics in materials with extended 5d-electron wave functions. Doping this spin-orbit Mott phase has since generated predictions of stabilizing states analogous to those found in doped strongly correlated 3d-electron Mott insulators such as the high temperature cuprate superconductors[4]. To date however, the role of Coulomb interactions in the doped $J_{eff}$=1/2 Mott phase remains contentious with no direct observations of correlated electronic phase behavior.

Two prototypical spin-orbit Mott materials are the *n*=1 and *n*=2 members of the iridate Ruddelsden-Popper series $Sr_{n+1}Ir_nO_{3n+1}$ [5,6]. Here the bilayer system $Sr_3Ir_2O_7$ (Sr-327) possesses a low temperature charge-gap of $E_g$=130 meV,[7] roughly reduced by a factor of four from the gap of its single layer cousin $Sr_2IrO_4$.[8] This reduced gap renders the Sr-327 system a fortuitous starting point for perturbing the spin-orbit Mott phase and exploring carrier-induced electronic phase behavior as the system is driven toward the metallic regime. To this end, in this work $Ru^{4+}$ ($4d^4$) ions are substituted onto the $Ir^{4+}$ ($5d^5$) sites of $Sr_3(Ir_{1-x}Ru_x)_2O_7$ with the known end point, $Sr_3Ru_2O_7$, possessing a Fermi liquid ground state in close proximity to a magnetic instability[9]. Our combined transport, magnetization, neutron scattering, and scanning-tunneling spectroscopy (STS) studies show that the Mott insulating state of $Sr_3Ir_2O_7$ is remarkably robust as the in-plane doped holes remain largely localized within a nanoscale phase separated ground state and only generate a metal-insulator transition (MIT) near the 2D percolation threshold. The resulting electronic phase diagram also reveals the surprising persistence of antiferromagnetic (AF) order deep into the metallic phase and suggests emergent itinerant magnetism at the interface between the AF-ordered spin-orbit Mott phase of $Sr_3Ir_2O_7$ and the nearly magnetic Fermi-liquid electronic phase of $Sr_3Ru_2O_7$.

## Results

### Electronic phase diagram and bulk electronic properties

The resulting electronic phase diagram determined via our combined transport, bulk magnetization, and neutron scattering measurements is plotted in Fig. 1 (a). The most prominent feature of the phase diagram is that the transition from the insulating ground state of $Sr_3Ir_2O_7$ to the low temperature metallic phase takes place only beyond the critical concentration of $x$=0.35. This suggests that the $Ir^{4+}$ ($5d^5$) valence is protected by the Mott gap that blocks the charge transfer of doped holes from the in-plane substituted $Ru^{4+}$ ($4d^4$) ions, a phenomenon of “Mott blocking”. The corresponding resistivity $\rho(T)$ is plotted as a function of temperature in Fig. 2 (a) for $Sr_3(Ir_{1-x}Ru_x)_2O_7$ concentrations spanning the phase diagram. Concentrations near the phase boundary also show a thermally driven MIT as illustrated in the inset of Fig. 2(a) for $x$=0.33 with $T_{MIT}$=135 K (see also Supplementary Fig. 1). As an initial window into the corresponding evolution of the magnetic order, the high temperature inflection in $\rho(T)$ in the $x$=0 parent compound is known to identify the onset of canted AF order at $T_{AF}$=280 K. This feature in $\rho(T)$ is gradually suppressed to lower temperatures upon Ru-doping, where the anomaly vanishes in the metallic regime.

Low temperature magnetoresistance (MR) data with the magnetic field applied perpendicular to the *ab*-plane are plotted in Fig. 2 (b). The negative magnetoresistance previously reported in the parent material[10] and indicative of suppressed spin fluctuations or magnetic domain scattering persists in lightly doped, insulating samples; however, as the system transitions into the metallic phase, the MR smoothly switches sign from negative to positive values that increase in magnitude with continued Ru-doping. This suggests that orbital (Lorentz force) effects begin to dominate across the MIT phase boundary as the carrier concentration is enhanced while fluctuation/domain effects from AF order are damped. Further illustrating this, bulk magnetization measurements of the in-plane susceptibility were performed on select samples, shown in Fig. 2 (c). As Ru is doped into Sr-327, the onset temperature of the net ferromagnetism, arising from the canted AF order and denoted via the irreversibility temperature ($T_{irr}$), is reduced. Close to the critical regime, the $x$=0.33 sample exhibiting a thermally driven MIT with $T_{MIT}$=135 K (Fig. 2 (a) inset) shows an onset of canted AF order at the same temperature. This suggests that near the MIT phase boundary the two transitions ($T_{MIT}$ and $T_{CAF}$) become coupled and that this coupling diminishes in lightly doped samples deeper within the insulating regime. Samples with Ru-doping $x$>0.33 show no irreversibility in magnetization, and concentrations with a metallic ground state show only local moment behavior within resolution. The only exception is that the highest doped sample with $x$=0.75 shows the reemergence of $T_{irr}$ at low temperature (Supplementary Fig. 2); however the origin of this may simply be an extrinsic perturbation of the nearby Fermi-liquid phase of $Sr_3Ru_2O_7$.

Through direct analogy with $Sr_3Ru_2O_7$,[11] Ru nominally enters the Sr-327 iridate lattice in the low spin state of $Ru^{4+}$ and subsequently introduces $S$=1 impurities into the $J_{eff}$=1/2 magnetic background. Unlike its single layer cousin $Sr_2IrO_4$,[12] the parent Sr-327 iridate shows no Curie-Weiss behavior up to 400 K[13]; but as Ru-ions are introduced into the lattice a paramagnetic upturn begins to build in the low temperature magnetization data for the lowest doping measured ($x$=0.13). Immediately upon doping Ru, the known low-temperature downturn in $\chi(T)$ in the parent system[10,14] (Fig. 2 (c)) rapidly vanishes and is replaced by a weak paramagnetic upturn. The resulting local moments, extracted via Curie-Weiss fits to the susceptibility, are plotted in Fig. 2 (d). For low Ru-dopant levels, the effective local moments extracted from each concentration track the expectation for contributions solely arising from local $S$=1 impurities which build continuously across the MIT. This suggests Ru-ions remain largely localized at low Ru-dopings within the insulating background of $Sr_3Ir_2O_7$ and that their survival into the metallic regime demonstrates robust correlation effects on either side of the MIT. For doping levels beyond $x$=0.5, the local moments are screened and smoothly connect to the high-temperature susceptibility of metallic $Sr_3Ru_2O_7$.[11]

**Neutron scattering measurements**

In order to more directly elucidate the evolution of the ordered antiferromagnetic phase across the MIT in this system, neutron scattering measurements were performed. The results plotted in Fig. 3 (a) show that, for insulating samples, the onset of long-range AF order coincides with the $T_{irr}$ determined via the magnetization curves in Fig. 2. Upon increased doping, however, the AF phase surprisingly survives across the MIT at the same **Q**–positions as the insulating phase,[10,15] and the resulting order parameters for metallic samples are plotted in Fig. 3 (b). From the limited number of magnetic peaks observable in our neutron measurements ((1, 0, $L$ ); $L$ = 1, 2, 3, 4), the spin structure remains consistent with that of the parent system across the MIT in the phase diagram, albeit the small degree of spin canting present in the insulating parent system is necessarily eliminated or strongly suppressed in the metallic regime. The persistent AF order remains long-range within resolution with a minimum correlation length $\xi_{min}\approx 200$ Å ( $\xi = 2\sqrt{2\ln(2)}\frac{1}{w}$ where $w$ is obtained by fits of radial scans to the form $I = I_0 + Ae^{-\frac{1}{2}\left(\frac{(x-c)}{w}\right)^2}$ ). Keeping a model of c-axis aligned moments across the MIT,[5] Fig. 3 (c) shows a nearly linear suppression of the AF moment in the lightly Ru-doped, insulating regime due to the dilution of ordered Ir-ions by localized Ru $S$=1 impurities, and deep in the metallic regime the ordered moment is quickly screened. In close proximity to the MIT phase boundary however, an anomalous enhancement in the ordered AF moment appears (Fig. 3 and Supplementary Fig. 3), suggesting the potential of induced ordering of $S$=1 moments from doped $4d^4$ electrons in this range or potentially a partial relaxation of the octahedral distortion resulting in enhanced magnetic exchange.

The intrinsic crystal structure of $Sr_3Ir_2O_7$ remains an active area of investigation with superlattice reflections violating the tetragonal space group *I4/mmm* reported in single crystal studies[14,16]. Previously, our neutron studies resolved high-temperature Bragg scattering[10] at positions forbidden by both the recently reported *I4/mmm*[15] and *Bbcb*[5] space groups. In order to clarify the origin of this high temperature superlattice, we also performed polarized neutron diffraction measurements with the results plotted in Fig. 3 (d). Radial scans through **Q**=(1,0,3) show that the (1, 0, *L*)-type superlattice reflections at 300 K appear only in the non-spin-flip channel with the neutron guide field applied parallel to **Q**. This demonstrates the structural origin of the superlattice and mandates a space group symmetry lower than *Bbcb*. The resulting high temperature (1, 0, *L*)-type peaks argue for oxygen octahedral tilting as well as in-plane rotation in this system. A tilt already necessarily exists for the c-axis aligned moments in the canted AF phase, and the strong spin-lattice coupling in perovskite iridates [17] supports the notion of an accompanying structural tilt. Such a tilt likely renders Sr-327 isostructural to $Ca_3Ru_2O_7$ (space group: $Bb2_1m$)[18]; however a full neutron data set and structural refinement have yet to be carried out.

**Scanning tunneling spectroscopy measurements**

In order to better understand the formation of the metallic phase, low temperature (4K) STS measurements were performed on two concentrations: Samples with *x*=0.35 in close proximity to the MIT, and samples deep within the metallic regime with *x*=0.5. Fig. 4 (a) shows the resulting topography of STS measurements exploring the local density-of-states (LDOS) in the *x*=0.35 concentration. Strong inhomogeneity across nanometer length scales in this sample is immediately apparent from the topography and reveals the coexistence of two distinct local environments whose representative tunneling spectra are plotted in Fig. 4 (c). Dark regions with low LDOS in the corresponding map show a fully gapped spectra paralleling that of the parent $Sr_3Ir_2O_7$ insulating phase[7] reproduced in Fig. 4 (f), while the bright regions reveal metallic regions with an enhanced LDOS. The striking *nanoscale* coexistence of both fully gapped and gapless, metallic regions in this sample demonstrates that the sample segregates into electronically distinct regions. The low temperature MIT phase line in Fig. 1 (a) therefore does not represent a thermodynamic phase transition but rather the percolation threshold of metallic puddles localized within a spin-orbit Mott phase.

In exploring the extent of this segregation between electronic phases or doped-carriers further, we performed STS measurements on the metallic *x*=0.5 concentration. These measurements reveal this sample to be globally gapless; however, the spectra also resolve a substantial degree of electronic inhomogeneity within this nominal metal, as illustrated by a representative topography in Fig. 4 (b). Correspondingly, the spectra plotted in Fig. 4 (d) again show two distinct shapes representing different local environments: one with suppressed V-shaped LDOS, and the second with enhanced LDOS and a spectrum that strongly resembles that of $Sr_3Ru_2O_7$.[19] To better illustrate this, a comparison with $Sr_3Ru_2O_7$ is provided as shown in Fig. 4 (e). The similarity between the hole-rich regions of the metallic *x*=0.5 sample and the pure bilayer ruthenate system is particularly striking, with the tunneling data resembling a thermally broadened version of a qualitatively similar electronic structure. This combined with the strong inhomogeneity of this metallic state indicates that even the fully metallic compounds continue to remain electronically segregated over nanometer length scales.

## Discussion

Our combined experimental results, viewed globally, paint a picture of a nanoscale, electronically phase separated ground state for in-plane carriers doped within a spin-orbit driven Mott phase, $Sr_3Ir_2O_7$ (Fig. 1). Since the meaning of “electronic phase separation” is rather subtle at the nanoscale in doped transition metal oxides, we define its use explicitly here simply as the observation of two different local environments with distinct electronic properties. This general scenario of nanoscale phase separation, either via the coexistence of distinct electronic phases or the direct segregation of holes, results in the stabilization of two different local environments and a percolating conduction network sensed by our earlier transport measurements. Bare charge accumulation into puddles of 1-2 nanometer length scales may not be energetically favorable due to unscreened long-range Coulomb interactions. Without knowing the effective screening length for the Coulomb interaction and the pinning potential for carriers, it is hard to quantify what the length scale should be in $Sr_3Ir_2O_7$. An alternative of phase separation into electrically neutral, yet electronically distinct, phases separated by a first order phase transition is instead a likely mechanism; however we are unable to differentiate this from the pure carrier segregation scenario.

Regardless of which scenario dominates, the carriers within metallic patches remain initially localized across ~1-2 nm length scales, and at the critical concentration where transport measurements show a MIT ($x$~0.35), this leads to the formation of metallic patches percolating within the fully gapped, spin-orbit Mott insulating background. At Ru-substitution levels below $x$=0.35, the thermally driven MIT is therefore the likely result of the expansion of these metallic puddles due to thermal shifts in their free energy relative to insulating host phase. Phase inhomogeneity continues deep into the metallic regime, where our STS data directly demonstrate nm-scale texture in metallic $Sr_3IrRuO_7$ comprised of two distinct regions: (1) Large LDOS regions with an electronic response mirroring the $4d^4$ electronic spectrum of isostructural $Sr_3Ru_2O_7$[19] and (2) Regions with V-shaped spectra with LDOS suppressed close to the Fermi energy.

Since their valence states are rather far from the Fermi-level, A-site doping in perovskite oxides is historically envisioned as controlling the filling of d-bands on the B-sites by donating their valence electrons to the entire system. The resulting doping mechanism gives rise to a rapid suppression of the Mott phase such as in A-site doped $Sr_2IrO_4$[20] and $Sr_3Ir_2O_7$.[21] Our B-site doping in Sr-327, however, reveals that holes nominally added via Ru-substitution remain localized within the $IrO_2$-planes until nearly 35% of the Ir 5d ions have been replaced, close to the classical 2D percolation threshold of 41%.[22] Even beyond this threshold at 50% replacement, hole-rich regions remain phase separated. Given that Ru-doping is nominally a strong perturbation to the weakly insulating ground state of Sr-327, this observation is striking and suggests that Coulomb interactions and correlation effects remain essential across the majority of the phase diagram of this system.

Our combined neutron scattering and STS data reveal that the AF ordered state that survives across the MIT has a spin-spin correlation length ($\xi$>200 Å) that spans across the phase separated puddles of gapped and metallic regions—revealing a globally AF ordered phase. Furthermore in concentrations doped close to the MIT, the recovery of the ordered AF moment to values nearly equaling that of the undoped parent $Sr_3Ir_2O_7$ rules out any trivial superposition of chemically distinct phases. A magnetically ordered, metallic state beyond the MIT is reminiscent of the phase diagrams of $(Ca_{1-x}Sr_x)_3Ru_2O_7$[23] and $Ca_{2-x}Sr_xRuO_4$[24]; however, from our current measurements of $Sr_3(Ir_{1-x}Ru_x)_2O_7$ the structural symmetry appears identical for concentrations spanning the MIT, suggesting that the critical point is not directly tied to a structural phase transition. AF metallic states have also been proposed in disordered and binary alloy Mott phases as an intermediate state prior to the onset of Anderson localization[25,26]. The global picture our data provide show that the physics here is more complex than that of a trivially diluted AF system with percolative transport. The percolating metallic network seemingly can be induced to order by the host AF matrix, which may explain why the ordered AF moment is actually enhanced near the region of maximum heterogeneous interface area at the MIT as well as why AF order survives across the percolation threshold where no infinite domain of the AF host persists. Local antiferromagnetism does however naively persist across the critical concentration and can continue to influence the metallic phase into the heavily Ru-doped regime. Eventually this gives way to a globally gapless AF phase in the $x$=0.5 sample.

We propose the following picture of magnetic interactions within this system: When they are dilute within the matrix, Ru-doped holes behave in a manner consistent with isolated ions in the $S$=1 low spin state giving rise to the local moment response; however, increasing the Ru-doping level increases the density of these isolated magnetic impurities, eventually nucleating clusters of metallic regions (resolved directly in our STS measurements). Within these metallic puddles, whose percolation generates the MIT, the local moment should be quenched at low temperatures in a Fermi-liquid ground state; however, these puddles may still be magnetically ordered due to proximity of local AF order in neighboring regions and a large spin susceptibility arising from their nested Fermi surface pockets. Such an instability is indeed known to be present along the $\mathbf{Q}=(\pi, \pi)$ in-plane wave vectors of $Sr_3Ru_2O_7$[27] where an enhanced density of states is nested at the Fermi level due to the $\sqrt{2}\times\sqrt{2}$ structural zone folding. In this regard, this suggests similarities to the thermally driven MIT in the prototypical Mott system $VO_2$, where percolating metallic puddles display significant correlation effects[28]. More broadly, the survival of an ordered magnetic moment into the metallic state of the system demonstrates that electron-electron correlations remain relevant across the MIT of this system and argues against the picture of $Sr_3Ir_2O_7$ as a trivial band-insulator simply driven by the zone-folding that occurs at the onset of AF order.

The evolution of AF order across the MIT in the phase diagram of this hole-doped spin-orbit Mott insulator demonstrates that a rich interplay can be realized at the boundary between a novel $J_{eff}$=1/2 insulator and a correlated metal. The localization of Ru-doped carriers into a phase separated ground state surprisingly parallels the strongly correlated phase behavior of 3d-transition metal oxide systems such as the B-site doped correlated manganites[29,30,31,32] and reveals that correlation physics can play a dominant role in the electronic phase formation of a doped spin-orbit Mott insulator. Our findings demonstrate that correlation effects felt by carriers introduced within in a 5d Mott phase remain robust enough to drive electron localization, a key ingredient in emergent phenomena such as high temperature superconductivity and enhanced ferroic behavior. This opens up a new frontier for exploring correlated electron phases within the presence of strong spin-orbit coupling effects inherent to a 5d-electron setting.

## Methods

**Materials and crystal growth:** The single crystals of $Sr_3(Ir_{1-x}Ru_x)_2O_7$ were grown by conventional flux methods similar to earlier reports[10,14] using a $SrCl_2$ flux. Crystals were grown in platinum crucibles using $IrO_2$ (99.98%, Alfa Aesar), $RuO_2$ (99.98%, Alfa Aesar), $SrCO_3$ (99.99%, Alfa Aesar), and anhydrous $SrCl_2$ (99.5%, Alfa Aesar) in a 2:3:15 molar ratio. Starting powders were partially sealed inside the crucible with a Pt lid and further contained inside alumina crucibles. Mixtures were heated up to 1380 °C, cooled to 850 °C at a rate of 3.5 °C/hr, and then furnace cooled to room temperature. The resulting boule was etched with deionized water and shiny, black $Sr_3(Ir_{1-x}Ru_x)_2O_7$ crystals with typical dimensions 2 mm x 2 mm x 0.1 mm were removed.

Ru concentrations were determined to match target values within ~2% via energy dispersive x-ray spectroscopy (EDS) measurements. EDS measurements were performed on numerous samples across different regions of samples from each growth batch, and measurements were also collected across different length scales to verify chemical homogeneity. Multiple crystals were tested from every batch, and from point to point on a given sample, we were able to resolve a Ru distribution homogenous within a central value +/- 1% (2% spread). The central value of Ru-concentrations between crystals from a single growth batch would vary no more than +/- 2% from a central value (4% spread). Error bars on the reported phase diagram in the main text reflect this uncertainty---in many cases they are within the symbol size. The actual crystals measured via transport and magnetization measurements and almost all of the crystals for the neutron measurements (only the *x*=0.20 and *x*=0.15 samples were not, although crystals from the same batch were) were first characterized via EDS measurements to determine/verify the precise Ru-content.

**X-ray diffraction measurements:** Single crystals from a single batch of each concentration were ground into a powder and measured via x-ray powder diffraction within a Bruker D2 Phaser diffractometer. X-ray powder diffraction and refinement revealed no impurity phases within instrument resolution (~2-3%). Lattice parameters and unit cell volumes were refined within the *I4/mmm* space group and showed both *a*- and *c*-axes that reduce continuously with increased Ru substitution (Supplementary Fig. 4)---as expected as the smaller $Ru^{4+}$ ions are introduced into the lattice. We note here that laboratory-based powder x-ray measurements typically lack the intensity to resolve the known orthorhombic superlattice reflections in this material so each concentration was instead refined within the tetragonal *I4/mmm* space group.

**Bulk property measurements:** Magnetotransport measurements were performed via standard four-wire measurements within a Quantum Design PPMS. Magnetization measurements were collected within a Quantum Design SQUID MPMS magnetometer.

**Metal-Insulator transition determined via resistivity:** The metal-insulator transition (MIT) depicted in the phase diagram of Fig. 1 (a) was determined via the temperature at which the slope of the sample's resistance versus temperature changed sign from $\delta R/\delta T>0$ to $\delta R/\delta T<0$ upon cooling. Data showing the MIT for all samples where an MIT was reported are shown in the inset of Fig. 2 (a) and in Supplementary Fig. 1.

**Bulk spin susceptibility:** Local moments were determined via Curie-Weiss fits to the form $1/\chi(T) = \Theta/C + T/C$, where $\Theta$ is the Weiss constant and *C* is the Curie constant (results plotted in Supplementary Fig. 2). Here $C = \frac{N_A}{3k_B}\mu_{eff}^2$ with $N_A$ as Avogadro's number and $k_B$ is Boltzman's constant. Fits render a negative Weiss constant for $x$<0.75 consistent with the observation of antiferromagnetic (AF) correlations. For $x$=0.75, low temperature susceptibility shows the reemergence of an irreversibility temperature ($T_{irr}$) below 20K. The reentrance of a net ferromagnetic signal in this heavily doped regime is likely the result of the fragility of the nearly magnetic ground state of $Sr_3Ru_2O_7$ where small levels of impurity substitution or pressure are known to stabilize magnetic order[11]. As similar effects are known to occur in $Sr_3Ru_2O_7$, for the purposes of our study we treat the ground state of $x$=0.75 as qualitatively similar to that of the ruthenate bilayer endpoint. We have not performed neutron diffraction measurements exploring the presence of antiferromagnetic order in this concentration.

**Neutron scattering measurements:** The unpolarized neutron diffraction experiments were performed on HB1-A triple axis spectrometer at the High Flux Isotope Reactor (HFIR) at Oak Ridge National Laboratory (ORNL) and at N-5 triple axis spectrometer at Canadian Neutron Beam Center, Chalk River Canada. For HB1-A the incident beam was monochromated by the **Q**=(0, 0, 2) reflection of a double-bounce pyrolitic-graphite (PG) monochromator with a fixed incident energy of $E_i$=14.65 meV, and a PG(002) analyzer crystal was used on the scattered side. Two PG filters were placed before the sample, and collimations of 40´-40´-40´-80´ were used before the monochromator, sample, analyzer, and detector respectively. Experiments on N5 were performed with a PG monochomator and $E_i$ = 14.5 meV and PG analyzer with one PG filter placed after the sample. Collimations of 30´-60´-33´-144´ were used before the monochromator, sample, analyzer, and detector respectively. The polarized neutron experiment was carried out on the BT7 triple-axis spectrometer at the NIST Center for Neutron Research using PG(002) monochromator, $^3$He polarizers, PG filters before and after the sample, radial collimation, and a position sensitive detector on the scattered side. A guide field allowed the magnetic field to be tuned along the scattering vector (horizontal field) and perpendicular to the scattering plane (vertical field) configuration. For all experiments, the crystals were aligned in the (H,0,L) scattering plane.

**Scanning tunneling spectroscopy measurements:** $Sr_3(Ir_{1-x}Ru_x)_2O_7$ single crystals were cleaved at ~77K in ultra-high vacuum before being directly transferred to the STM head held at 4K. From previous data, cleaving at low temperatures is critical for obtaining flat clean samples. Tips were prepared by annealing etched W-tips in vacuum and then checking the quality on metallic (copper single crystal) surfaces. The quality of all tips used in this study was checked by imaging standing waves on Cu and performing spectroscopy. The tips thus prepared showed atomic resolution on the iridate samples and were stable, allowing us to obtain high quality *dI/dV* maps.

**Measurement statistics:**

Transport measurements: Each batch with a unique Ru-concentration was tested at a minimum of three times (in most cases more) on different crystals. The resulting transport behavior shown in the paper Fig. 2 reproduced within the errors shown within the phase diagram of Fig. 1. The most sensitive batches were those with Ru-contents close to the sharp MIT phase boundary with $x$=0.35 where the same slight deviations in Ru content resulted in a larger sample-to-sample variation in $T_{MIT}$. The resulting uncertainty is encompassed by the horizontal error bars in the main text's Fig. 1 (a), often within the symbol size.

Neutron measurements: Every data point for $T_{AF}$ on the phase diagram in Fig. 1 (a) represents a measurement on one unique sample. The anomalous regions such as those near the phase boundary with an enhanced moment ($x$=0.33-0.35) and deep within the metallic regime ($x$=0.5) were checked with additional experiments on additional samples grown in different batches. The magnetic behavior reproduced in both instances, and those data points are not shown.

Magnetization measurements: For our magnetization measurements, only a few select samples, well characterized by transport and EDS, were chosen for measurement. Only one measurement was taken for each concentration reported.

Scanning Tunneling Spectroscopy measurements: For each doping, at least three different tips and samples were studied. The spectral shapes for any given doping were consistent and repeatable. We checked the tip height dependence for the insulating regions of the parent compound and the $x$=0.35 compound and found no obvious changes in spectral shape with height. We have measured three samples at the $x$=0.35 concentration and found the reported nanoscale phase coexistence completely reproducible. Three samples with $x$=0.5 have also been measured and with completely reproducible spectra and surfaces.

**Acknowledgements:**

The work at Boston College was supported by NSF CAREER-Award DMR-1056625 (S.D.W), DOE DE-SC0002554 (Z.W.), and NSF DMR-1305647 (V.M.). The work at the ORNLs High Flux Isotope Reactor was sponsored by the Scientific User Facilities Division, Office of Basic Energy Sciences, US DOE. The identification of any commercial product or trade name does not imply endorsement or recommendation by the National Institute of Standards and Technology.

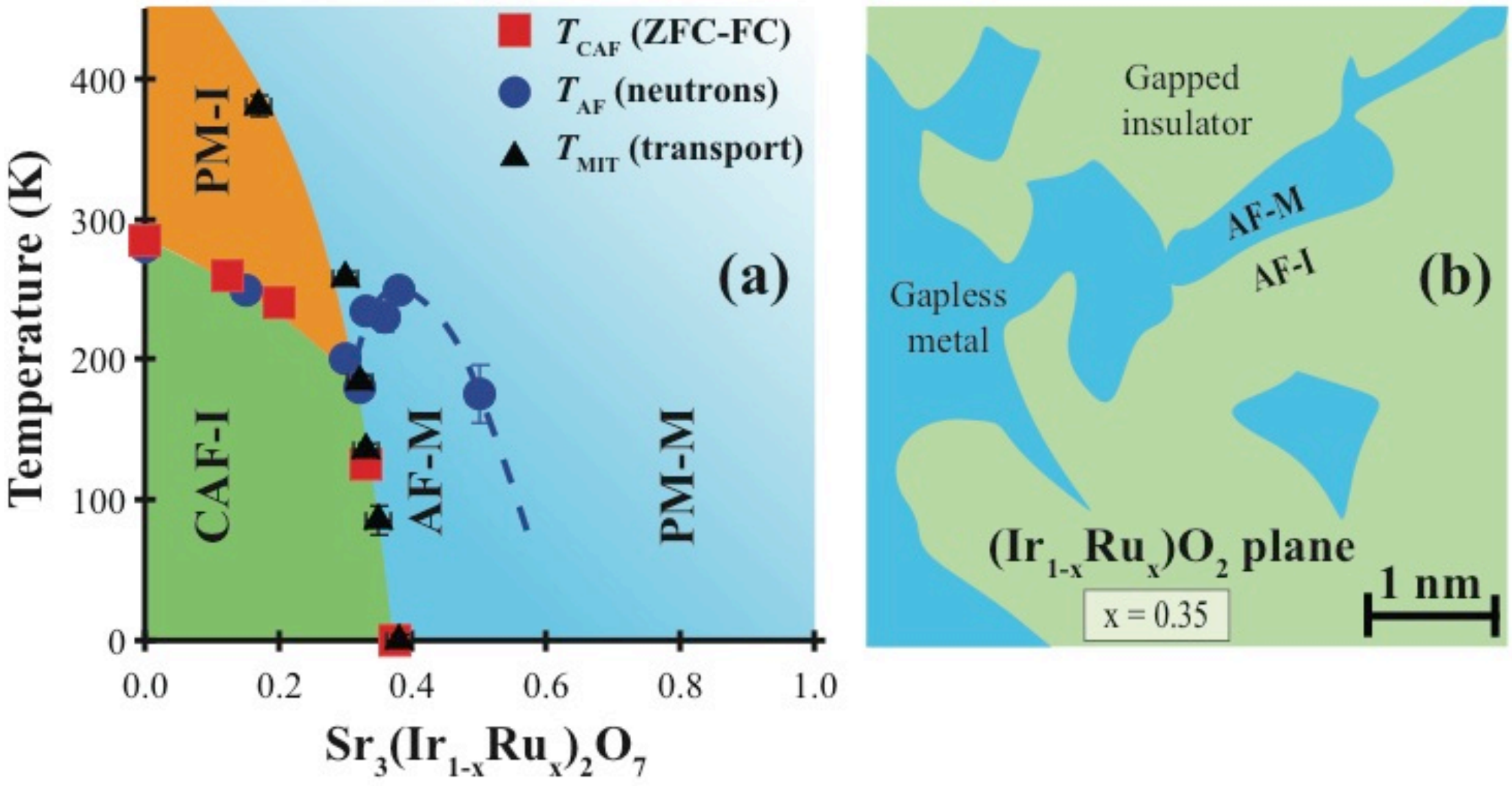

**Figure 1: Electronic phase diagram of $Sr_3(Ir_{1-x}Ru_x)_2O_7$** (a) Evolution of electronic phases of $Sr_3(Ir_{1-x}Ru_x)_2O_7$ as a function of Ru concentration. CAF-I denotes the insulating canted antiferromagnetic phase, PM-I denotes the paramagnetic insulating phase, AF-M denotes the AF ordered metallic state, and PM-M indicates the paramagnetic metallic regime. Squares indicate the onset of canted AF order determined with bulk susceptibility measurements, circles denote the onset of AF order as observed via neutron diffraction measurements, and triangles indicate the transition temperatures for thermally driven MITs near the phase boundary. (b) Illustration of the basal-plane showing phase separated metallic puddles near the percolative threshold, which nucleate within the spin-orbit Mott insulating background of $Sr_3Ir_2O_7$. Error bars in all plots represent one standard deviation.

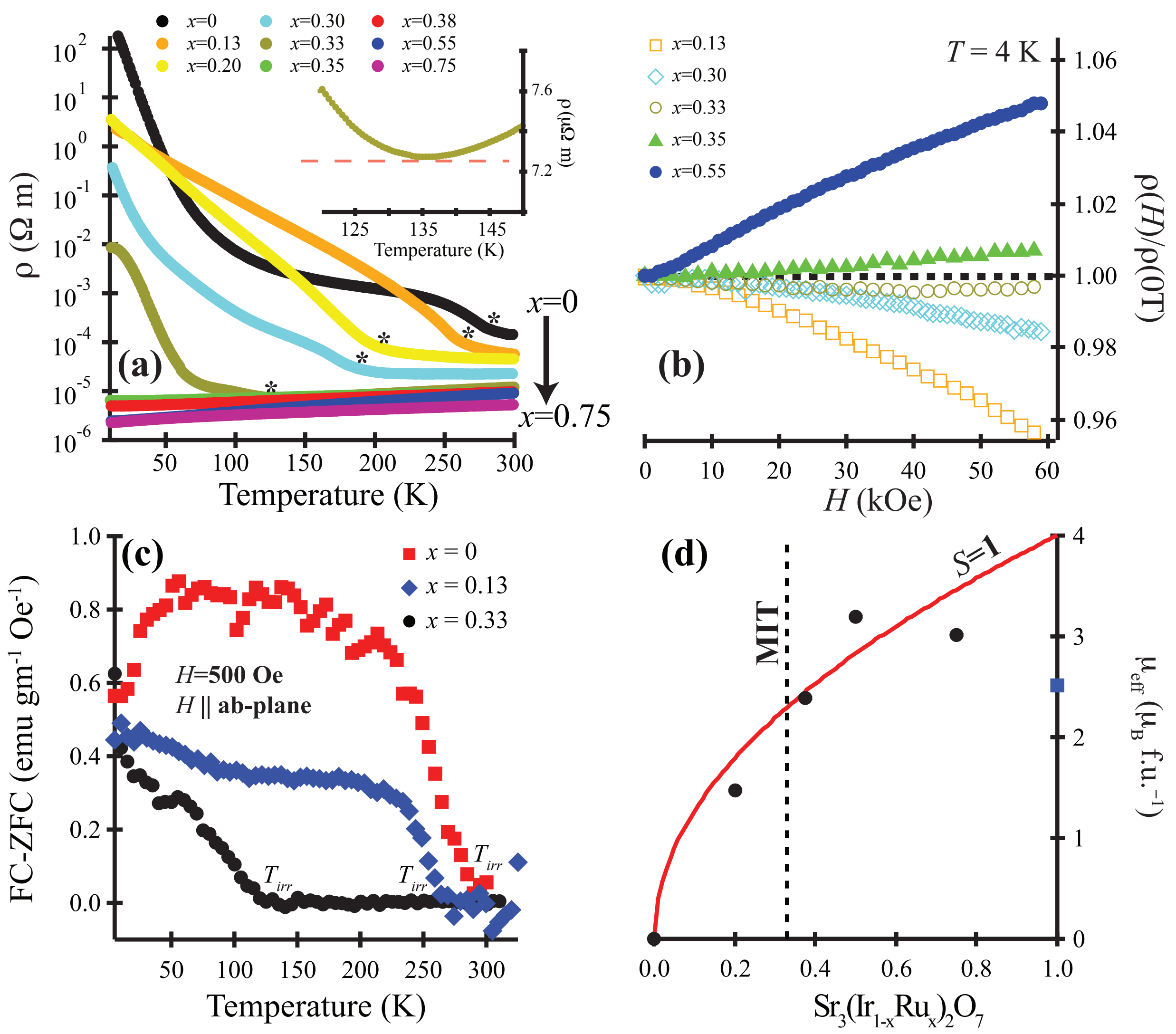

**Figure 2: Bulk transport and magnetization measurements of $Sr_3(Ir_{1-x}Ru_x)_2O_7$** (a) Resistivity plotted as a function of temperature for Ru concentrations spanning the MIT. Inset shows thermally driven transition at $T_{MIT}$=135 K for *x*=0.33. (b) 4 K magnetoresistance plotted as a function of applied field for Ru concentrations spanning the MIT. (c) Field cooled (FC) minus zero-field cooled (ZFC) magnetization as a function of temperature for select Ru dopings. (d) Local moments extracted from Curie-Weiss fits plotted as a function of Ru concentration. Solid line denotes the expected full moment value for *S*=1 impurities. Blue square shows data taken from Ikeda et al. (Ref. 11). 1 emu $g^{-1}$ $Oe^{-1}$ = 4π x $10^{-3}$ $m^3$ $kg^{-1}$. Error bars in all plots represent one standard deviation.

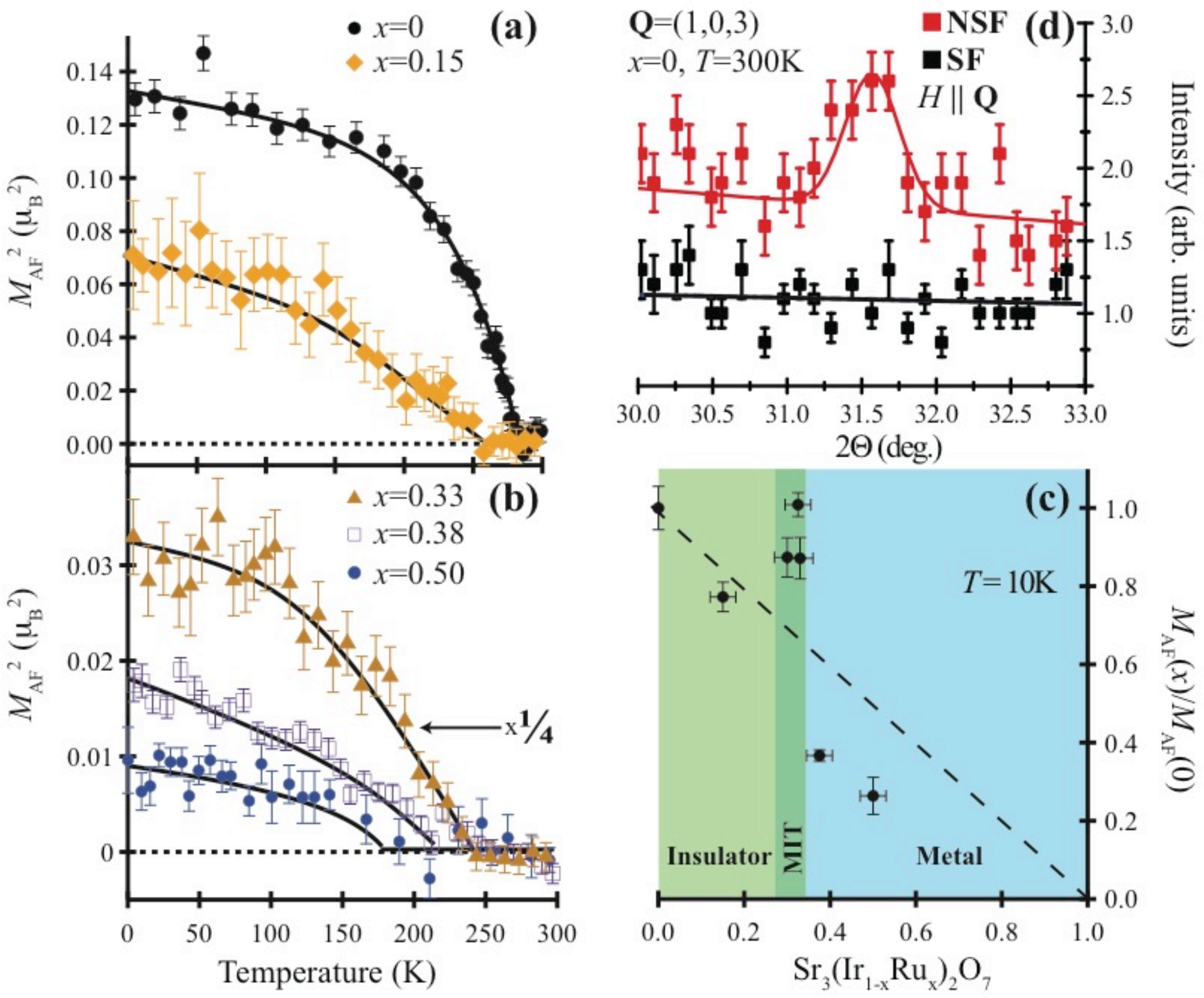

**Figure 3: Neutron scattering characterization of $Sr_3(Ir_{1-x}Ru_x)_2O_7$.** (a) Temperature evolution of the square of the antiferromagnetic order parameter ($M_{AF}^2$(T)) for fully insulating $x$ = 0 (from Ref. 10) and $x$ = 0.15. (b) $M_{AF}^2(T)$ for metallic $x$=0.5 and $x$=0.38 samples as well as for $x$=0.33 at the phase boundary. The data for $x$=0.33 have been scaled by ¼ for clarity. (c) Ordered AF moments for $Sr_3(Ir_{1-x}Ru_x)_2O_7$ scaled relative to the parent $Sr_3Ir_2O_7$ insulator. Shaded areas denote boundaries between insulating, metallic, and critical MIT regimes. (d) Polarized neutron diffraction measurements of $Sr_3Ir_2O_7$ showing radial scans through the **Q**=(1,0,3) superlattice peak at 300K in both spin-flip (SF) and non-spin-flip (NSF) channels. The magnetic guide field was applied parallel to the momentum transfer **Q**. Error bars in all plots represent one standard deviation.

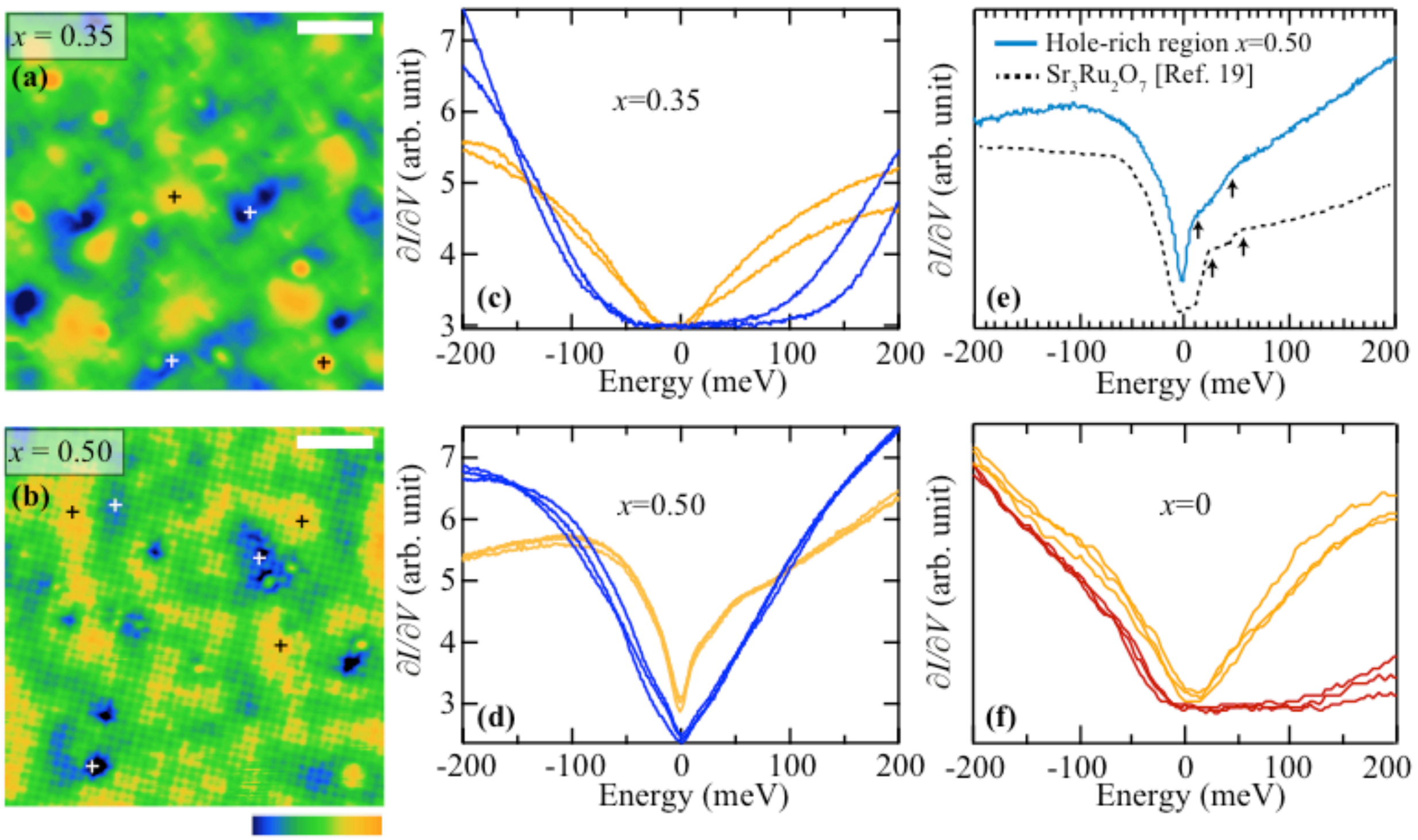

**Figure. 4: STS measurements of $Sr_3(Ir_{1-x}Ru_x)_2O_7$.** (a) Topography at a bias voltage of -100 mV for the *x*=0.5 concentration near the MIT. (b) Topography at a bias voltage of -100 mV for the *x*=0.5 metallic concentration. Intensity scales for topography in panels (a) and (b) are unique to each plot and their relative variation is shown via the color bar below the panels. White bars in each plot denote the length scale of 3 nm. Color bar denotes the relative scale of LDOS ranging from low (blue) to high (orange) values. (c) Tunneling spectra collected in gapped and gapless phase separated regions denoted by blue and yellow curves collected at white and black crosses in the corresponding map (d) Spectra for *x*=0.5 within two regions denoted by yellow and blue curves collected at black and white crosses respectively in the topography. (e) High resolution tunneling data collected within a bright region of the *x*=0.5 sample in panel (b). Dashed line is low temperature STS data for $Sr_3Ru_2O_7$ reproduced from Ref. 19. (f) Spectra collected for $Sr_3Ir_2O_7$ in regions with enhanced LDOS due to oxygen defects (yellow curves) and spectra collected away from defects showing the full charge gap (red curves). We note here that the inhomogeneity observed within the parent insulating system stems from relatively rare regions of oxygen defects and that the majority of the surface showed fully gapped behavior (red curves), whereas for the *x*=0.35 and *x*=0.5 systems the entirety of the samples showed strong electronic inhomogeneity across nanometer length scales.

Supplementary Figures:

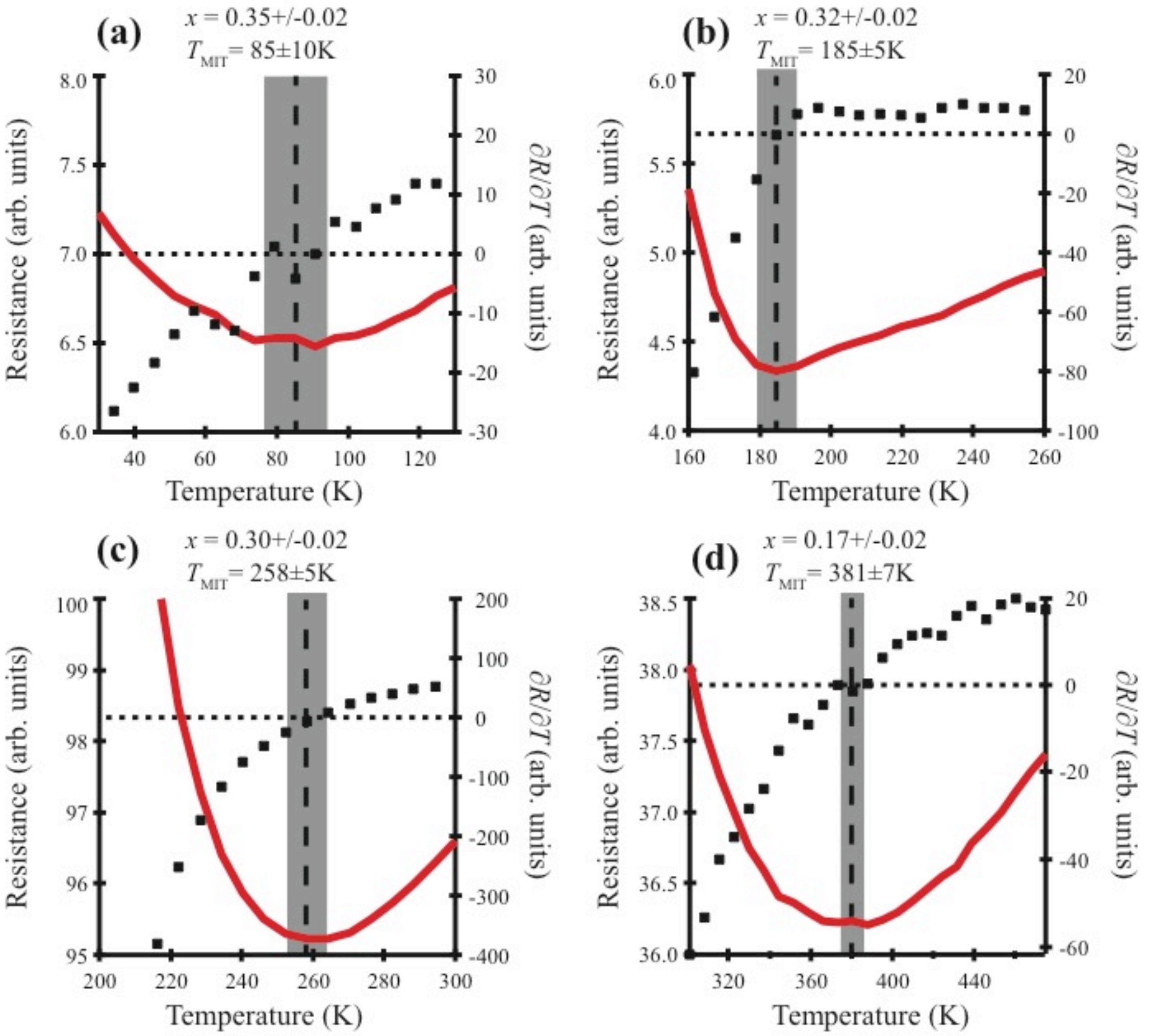

**Supplementary Figure 1:** Resistance (red lines) and its first derivative *δR/δT* plotted (black squares) plotted as a function of temperature for $Sr_3(Ir_{1-x}Ru_x)_2O_7$ samples with (a) *x*=0.35, (b) *x*=0.32, (c) *x*=0.3, and (d) *x*=0.17 respectively. Vertical dashed lines show the temperature $T_{MIT}$ plotted in the electronic phase diagram of Fig. 1 (a) in the paper's main text. Shaded grey region shows the uncertainty in determining $T_{MIT}$. Ru x-concentrations were measured for each sample with the corresponding uncertainty displayed in each panel. Resistance and *δR/δT* are plotted as raw data in arbitrary units without geometric conversion to resistivity.

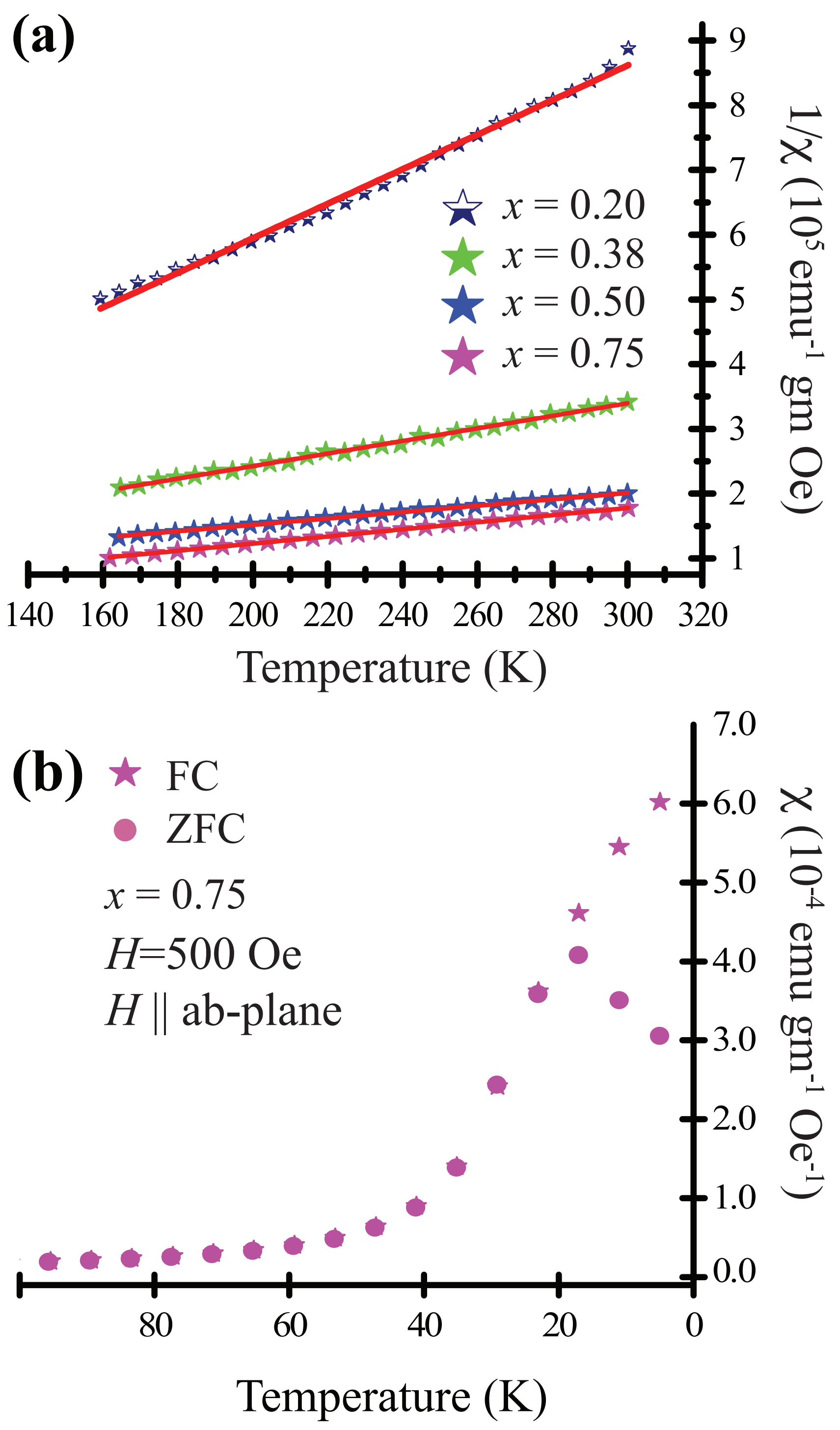

**Supplementary Figure 2:** (a) $\chi^{-1}$ as a function plotted as a function of temperature. A 500 Oe field was applied parallel to the ab-plane for all concentrations with the exception of *x*=0.2 where a 1 T field was applied. Solid lines are fits to Curie-Weiss behavior. (b) Zero-field cooled (ZFC) and field cooled (FC) magnetization data collected for the *x*=0.75 concentration. 1 emu $g^{-1}$ $Oe^{-1}$ = $4\pi$ x $10^{-3}$ $m^3$/kg.

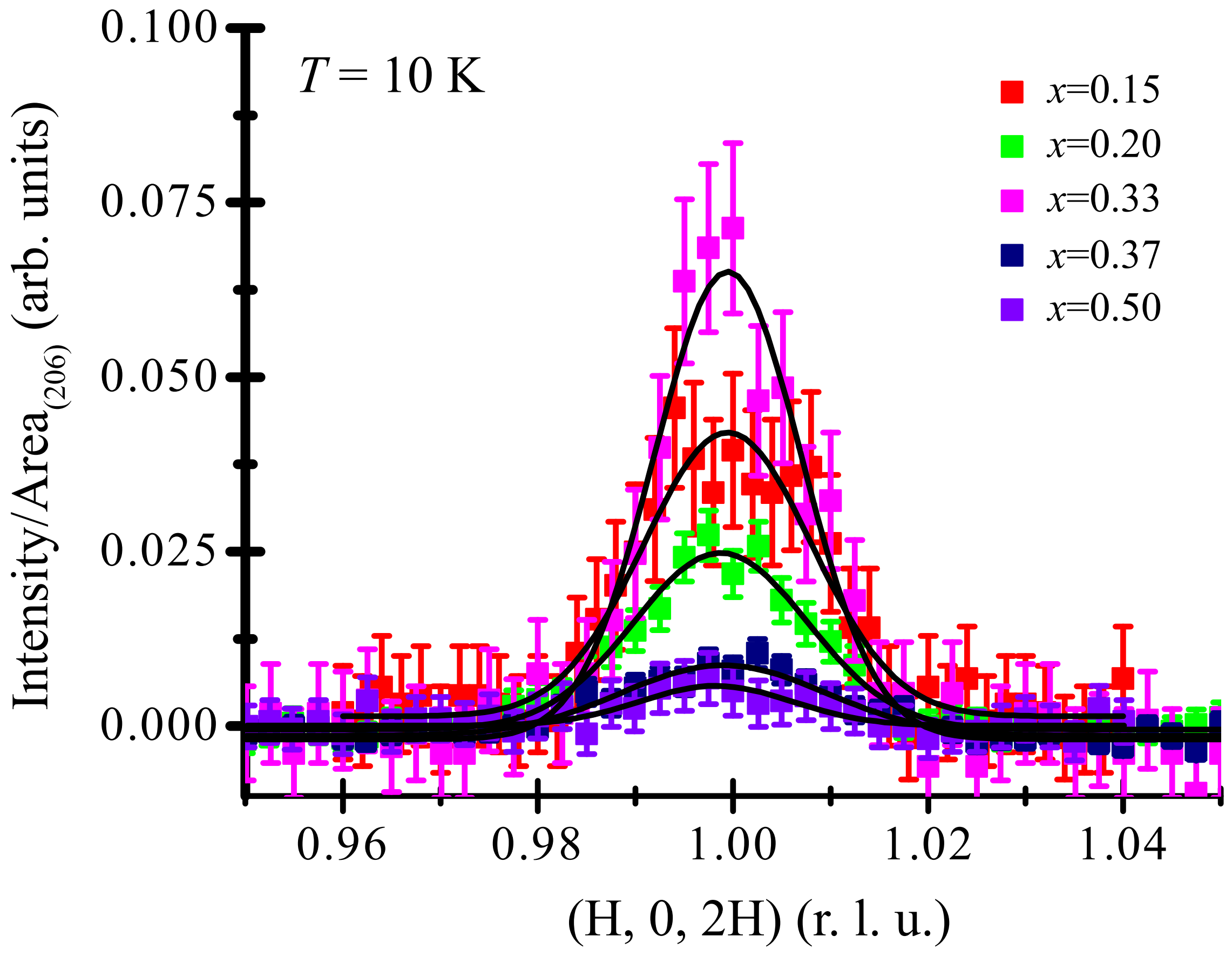

**Supplementary Figure 3:** Radial scans through the antiferromagnetic Bragg reflection **Q**=(1,0,2) for select concentrations of $Sr_3(Ir_{1-x}Ru_x)_2O_7$. Intensity for each sample has been divided by the integrated area of the sample's corresponding **Q**=(2,0,6) nuclear Bragg reflection. Before correcting for minor changes in absorption and extinction between Ru-concentrations this plot provides a rough illustration of the moment evolution as a function of *x*. Solid lines are Gaussian fits to the data. Error bars are one standard deviation. An additional sample, not plotted here, with nominal *x*=0.32 was measured with a normalized scattering intensity that saturates at 0.3 (off the scale) and the corresponding AF moment plotted in Fig. 3 (c) of the main text.

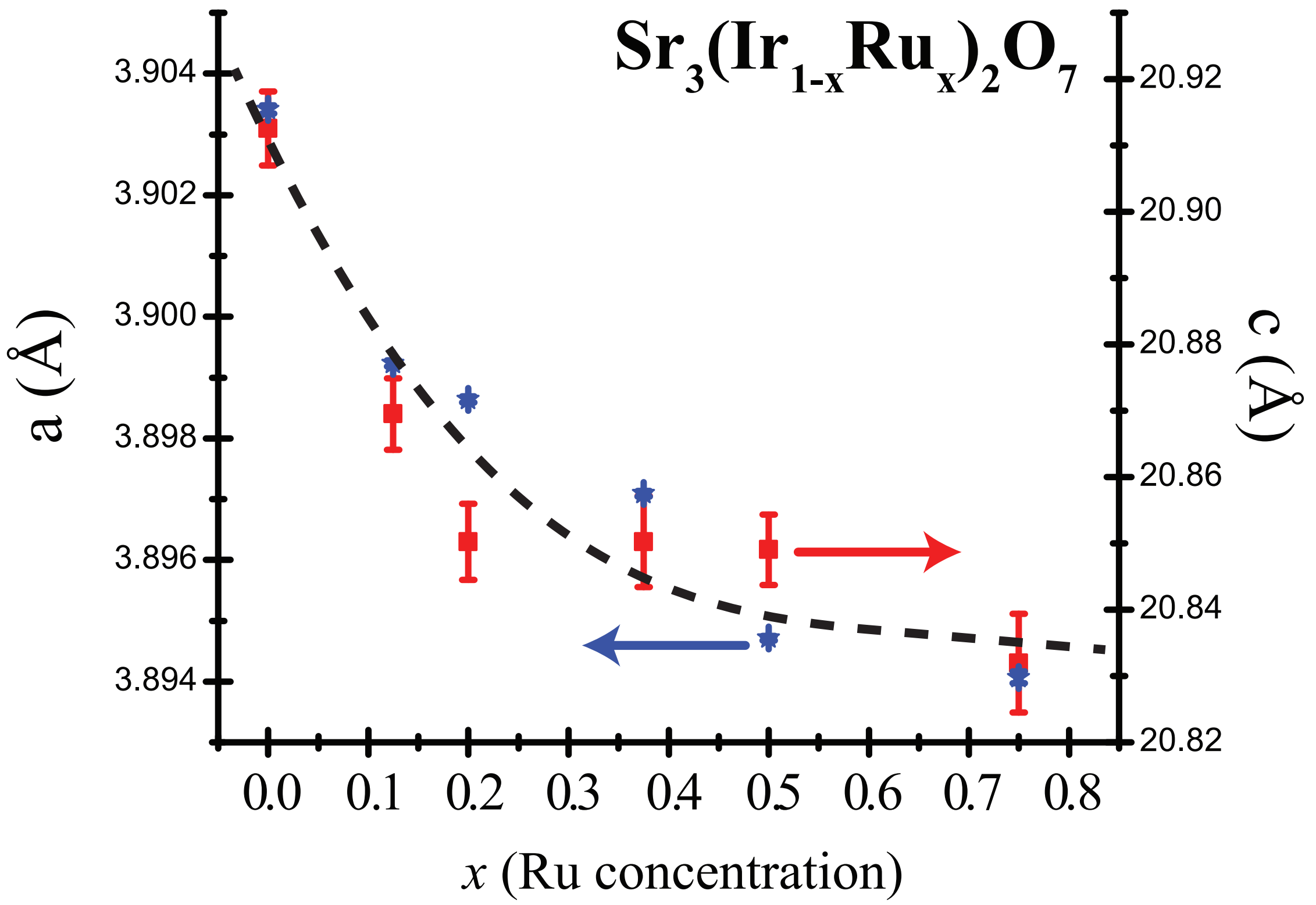

**Supplementary Figure 4:** Lattice parameters determined via powder x-ray diffraction on crushed crystals plotted as a function of Ru-doping for $Sr_3(Ir_{1-x}Ru_x)_2O_7$ at 300K. Data was refine using the FullProf Rietveld refinement program[S1]. Dashed line is a guide to the eye. While the reduction in lattice constants is monotonic, it is not linear and seemingly maps the nonlinear contraction previously observed in $Sr_2(Ir_{1-x}Ru_x)O_4$.[S2] Error bars are one standard deviation.

**References:**

[1] B. J. Kim, H. Ohsumi, T. Komesu, S. Sakai, T. Morita, H. Takagi, and T. Arima, Phase-Sensitive Observation of a Spin-Orbital Mott State in $Sr_2IrO_4$, Science 323, 1329-1332 (2009).

[2] B. J. Kim, Hosub Jin, S. J. Moon, J.-Y. Kim, B.-G. Park, C. S. Leem, Jaejun Yu, T.W. Noh, C. Kim, S.-J. Oh, J.-H. Park, V. Durairaj, G. Cao, and E. Rotenberg, Novel $J_{eff}=1/2$ Mott State Induced by Relativistic Spin-Orbit Coupling in $Sr_2IrO_4$, Phys. Rev. Lett. 101, 076402 (2008).

[3] Q. Wang, Y. Cao, J. A. Waugh, S. R. Park, T. F. Qi, O. B. Korneta, G. Cao, and D. S. Dessau, Dimensionality-controlled Mott transition and correlation effects in single-layer and bilayer perovskite iridates, Phys. Rev. B 87, 245109 (2013).

[4] F. Wang, T. Senthil, Twisted Hubbard Model for $Sr_2IrO_4$: Magnetism and Possible High Temperature Superconductivity, Phys. Rev. Lett. 106, 136402 (2011).

[5] J. W. Kim, Y. Choi, Jungho Kim, J. F. Mitchell, G. Jackeli, M. Daghofer, J. van den Brink, G. Khaliullin, and B. J. Kim, Dimensionality Driven Spin-Flop Transition in Layered Iridates, Phys. Rev. Lett. 109, 037204 (2012).

[6] S. J. Moon, H. Jin, K. W. Kim, W. S. Choi, Y. S. Lee, J. Yu, G. Cao,A. Sumi, H. Funakubo, C. Bernhard, and T. W. Noh, Dimensionality-Controlled Insulator-Metal Transition and Correlated Metallic State in 5*d*Transition Metal Oxides $Sr_{n+1}Ir_nO_{3n+1}$ ($n$=1, 2, and ∞), Phys. Rev. Lett. 101, 226402 (2008).

[7] Yoshinori Okada, W. Zhou, D. Walkup, Chetan Dhital, Ziqiang Wang, Stephen D. Wilson, and V. Madhavan, Imaging the evolution of metallic states in a spin-orbit interaction driven correlated iridate, Nature Materials 12, 707-713 (2013).

[8] Jixia Dai, Eduardo Calleja, Gang Cao, and Kyle McElroy. Local density of states study of a spin-orbit-coupling induced Mott insulator $Sr_2IrO_4$. Preprint at http://arXiv:1303.3688 (2013).

[9] S. A. Grigera, R. S. Perry, A. J. Schofield, M. Chiao, S. R. Julian, G. G. Lonzarich, S. I. Ikeda, Y. Maeno, A. J. Millis, and A. P. Mackenzie, Magnetic Field-Tuned Quantum Criticality in the Metallic Ruthenate $Sr_3Ru_2O_7$, Science 294, 329-332 (2001).

[10] Chetan Dhital, Sovit Khadka, Z. Yamani, Clarina de la Cruz, T. C. Hogan, S. M. Disseler, Mani Pokharel, K. C. Lukas,Wei Tian, C. P. Opeil, Ziqiang Wang, and Stephen D. Wilson, Spin ordering and electronic texture in the bilayer iridate $Sr_3Ir_2O_7$**,** Phys. Rev. B 86, 100401(R) (2012).

[11] Shin-Ichi Ikeda, Yoshiteru Maeno, Satoru Nakatsuji, Masashi Kosaka, and Yoshiya Uwatoko, Ground state in $Sr_3Ru_2O_7$: Fermi liquid close to a ferromagnetic instability, Phys. Rev. B 62, 6089(R) (2000).

[12] N. S. Kini, A. M. Strydom, H S Jeevan, C Geibel, and S Ramakrishnan, Transport and thermal properties of weakly ferromagnetic $Sr_2IrO_4$, J. Phys.: Condens. Matter 18, 8205 (2006).

[13] I Nagai, Y Yoshida, S I Ikeda, H Matsuhata, H Kito, and M Kosaka, Canted antiferromagnetic ground state in $Sr_3Ir_2O_7$, J. Phys.: Condens. Matter 19, 136214 (2007).

[14] G. Cao, Y. Xin, C. S. Alexander, J. E. Crow, P. Schlottmann, M. K. Crawford, R. L. Harlow, and W. Marshall, Anomalous magnetic and transport behavior in the magnetic insulator $Sr_3Ir_2O_7$, Phys. Rev. B 66, 214412 (2002).

[15] S. Boseggia, R. Springell, H. C.Walker, A. T. Boothroyd, D. Prabhakaran, D. Wermeille, L. Bouchenoire, S. P. Collins, and D. F. McMorrow, Antiferromagnetic order and domains in $Sr_3Ir_2O_7$ probed by x-ray resonant scattering, Phys. Rev. B 85, 184432 (2012).
[16] Hirofumi Matsuhataa,_Ichiro Nagaia, Yoshiyuki Yoshidaa, Sigeo Haraa, Shin-ichi Ikedaa, and Naoki Shirakawa, Crystal structure of $Sr_3Ir_2O_7$ investigated by transmission electron microscopy, Journal of Solid State Chemistry 177, 3776 (2004).
[17] G. Jackeli and G. Khaliullin, Mott Insulators in the Strong Spin-Orbit Coupling Limit: From Heisenberg to a Quantum Compass and Kitaev Models, Phys. Rev. Lett. 102, 017205 (2009).
[18] Yoshiyuki Yoshida, Shin-Ichi Ikeda, Hirofumi Matsuhata, Naoki Shirakawa, C. H. Lee, and Susumu Katano, Crystal and magnetic structure of $Ca_3Ru_2O_7$, Phys. Rev. B 72, 054412 (2005).
[19] K. Iwaya, S. Satow, T. Hanaguri, N. Shannon, Y. Yoshida, S. I. Ikeda, J. P. He, Y. Kaneko, Y. Tokura, T. Yamada, and H. Takagi, Local Tunneling Spectroscopy across a Metamagnetic Critical Point in the Bilayer Ruthenate $Sr_3Ru_2O_7$, Phys. Rev. Lett. 99, 057208 (2007).
[20] M. Ge, T. F. Qi, O. B. Korneta, D. E. De Long, P. Schlottmann, W. P. Crummett, and G. Cao, Phys. Rev. B 84, Lattice-driven magnetoresistivity and metal-insulator transition in single-layered iridates, 100402(R) (2011).
[21] L. Li, P. P. Kong, T. F. Qi, C. Q. Jin, S. J. Yuan, L. E. DeLong, P. Schlottmann, and G. Cao, Tuning the $J_{eff}$=1/2 insulating state via electron doping and pressure in the double-layered iridate $Sr_3Ir_2O_7$, Phys. Rev. B 87, 235127 (2013).
[22] Vinod K. S. Shante and Scott Kirkpatrick, An introduction to percolation theory, Advances in Physics 20, 325-357 (1971).
[23] Zhe Qu et al., Unusual heavy-mass nearly ferromagnetic state with a surprisingly large Wilson ratio in the double layered ruthenates $(Sr_{1-x}Ca_x)_3Ru_2O_7$, Phys. Rev. B 78, 180407(R) (2008).
[24] S. Nakatsuji et al., Heavy-Mass Fermi Liquid near a Ferromagnetic Instability in Layered Ruthenates, Phys. Rev. Lett. 90, 137202 (2003).
[25] N. Paris, A. Baldwin, and R. T. Scalettar, Mott and band-insulator transitions in the binary-alloy Hubbard model: Exact diagonalization and determinant quantum Monte Carlo simulations, Phys. Rev. B 75, 165113 (2007).
[26] Dariush Heidarian and Nandini Trivedi, Inhomogeneous Metallic Phase in a Disordered Mott Insulator in Two Dimensions, Phys. Rev. Lett. 93, 126401 (2004).
[27] A. Tamai, M. P. Allan, J. F. Mercure, W. Meevasana, R. Dunkel, D. H. Lu, R. S. Perry, A. P. Mackenzie, D. J. Singh, Z.-X. Shen, and F. Baumberger, Fermi Surface and van Hove Singularities in the Itinerant Metamagnet $Sr_3Ru_2O_7$, Phys. Rev. Lett. 101, 026407 (2008).
[28] M. M. Qazilbash, M. Brehm, Byung-Gyu Chae, P.-C. Ho, G. O. Andreev, Bong-Jun Kim, Sun Jin Yun, A. V. Balatsky, M. B. Maple, F. Keilmann, Hyun-Tak Kim, and D. N. Basov, Mott Transition in $VO_2$ Revealed by Infrared Spectroscopy and Nano-Imaging, Science 318, 1750-1753 (2007).
[29] M. Uehara, S. Mori, C. H. Chen, amd S.-W. Cheong, Percolative phase separation underlies colossal magnetoresistance in mixed-valent manganites, Nature 399, 560-563 (1999).
[30] Elbio Dagotto, Takashi Hotta, and Adriana Moreo, Colossal Magentoresistance Materials: The Key Role of Phase Separation, Physics Reports 344, 1-153 (2001).

[31] Kalpataru Pradhan, Anamitra Mukherjee and Pinaki Majumdar, Exploiting B-site disorder for phase control in the manganites, Euro. Phys. Lett. 84, 37007 (2008).
[32] A. Machida, Y. Moritomo, K. Ohoyama, T. Katsufuji, A. Nakamura, Phase separation and ferromagnetic transition in B-site substituted $Nd_{1/2}Ca_{1/2}MnO_3$, Phys. Rev. B 65, 064435 (2002).

Supplementary References:
[S1] J. Rodriguez-Carvajal, Recent Advances in Magnetic Structure Determination by Neutron Powder Diffraction, Physica B192, 55 (1993).

[S2] R. J. Cava, B. Batlogg, K. Kiyono, H. Takagi, J. J. Krajewski, W. F. Peck, Jr., L. W. Rupp, Jr., and C. H. Chen, Localized-to-itinerant electron transition in $Sr_2Ir_{1-x}Ru_xO_4$, Phys. Rev. B 49, 11890 (1994).